\documentclass[11pt,a4paper]{article}
\usepackage{amstex,amsthm,amssymb,epsfig}
\usepackage{cite,indentfirst}

\def\C|{{\mathbb C} \,}
\def\B|{{\mathbb B} \,}
\def\S|{{\mathbb S} \,}
\def\G|{{\mathbb G} \,}
\def\N|{{\mathbb N} \,}
\def\F|{{\mathbb F} \,}
\def\K|{{\mathbb K} \,}

\def\sg{\sigma}
\def\sgm{\sigma^-}
\def\sgp{\sigma^+}
\def\sgz{\sigma^z}

\def\Res{\operatorname{Res}}
\def\det{\operatorname{det}}

\newcommand{\bra}[1]{\langle\,#1\,|}
\newcommand{\ket}[1]{|\,#1\,\rangle}

\def\partialsur#1{\frac{\partial}{\partial#1}}

\newtheorem{theorem}{Theorem}[section]

   {
   
   }

\renewcommand{\theequation}{\thesection.\arabic{equation}}

\def\beqa{\begin{eqnarray}}
\def\eeqa{\end{eqnarray}}
\def\ba{\begin{array}}
\def\ea{\end{array}}
\def\r{\rangle}

\def\a{\alpha}
\def\b{\beta}

\def\e{\epsilon}

\def\th{\operatorname{th}}
\def\eps{\varepsilon}
\def\la{\lambda}
\def\s{\sigma}
\def\sul{\sum\limits}
\def\pl{\prod\limits}
\def\lt({\left(}
\def\rt){\right)}
\def\pd #1{\frac{\partial}{\partial #1}}

\def\th{\vartheta}

\begin{document}

\begin{titlepage}
\begin{flushright}
LPENSL-TH-15/99\\
\end{flushright}
\par \vskip .1in \noindent

\begin{center}
{\LARGE Correlation functions of the XXZ Heisenberg spin-$1 \over 2$ chain\\ 
in a magnetic field}\\
\end{center}
  \par \vskip .3in \noindent

\begin{center}

      {\bf N. KITANINE$^{*}$,  J. M. MAILLET, V. TERRAS}
  \par \vskip .1in \noindent

{\sl  Laboratoire de Physique $^{**}$\\
Groupe de Physique Th\'eorique\\
       ENS Lyon, 46 all\'ee d'Italie 69364 Lyon CEDEX 07
       France}\\[0.6in]
\end{center}

\par \vskip .10in
\begin{center}
{\bf Abstract}\\
\end{center}

\begin{quote}
Using the algebraic Bethe ansatz method, and the solution of the quantum 
inverse scattering problem for local spins, we obtain multiple integral representations of the $n$-point correlation functions of the XXZ 
Heisenberg spin-$1 \over 2$ chain in a constant magnetic field. For zero magnetic field, this result agrees, in both the massless and massive (anti-ferromagnetic) regimes, with the one obtained from the q-deformed KZ equations (massless regime) and the representation theory of the quantum 
affine algebra ${\cal U}_q (\hat{sl}_2)$ together with the corner transfer matrix approach (massive regime).
\end{quote}

\begin{flushleft}
\rule{5.1 in}{.007 in}\\
$^{*}$ {\small On leave of absence from the St Petersburg branch of the Steklov Mathematical Institute, Fontanka 27, St Petersburg 191011, Russia.}\\
$^{**}${\small UMR 5672 du CNRS, associ\'ee \`a  l'Ecole
Normale Sup\'erieure de Lyon.}\\
{\small This work is supported by CNRS (France), the EC-TMR contract FMRX-CT96-0012, MAE fellowship 96/9804 and MENRT (France) fellowship 
AC 97-2-00119.}\\
{\small email: Nicolai.Kitanine\symbol{'100}ens-lyon.fr, Jean-Michel.Maillet\symbol{'100}ens-lyon.fr, Veronique.Terras\symbol{'100}ens-lyon.fr}\\[0.2 in]

July 1999
\end{flushleft}

\end{titlepage}


\section{Introduction}
\setcounter{equation}{0}
\label{sect:intro}

The XXZ Heisenberg spin $\frac{1}{2}$ finite chain of length $M$~\cite{Hei28,Bet31}, in a constant magnetic field parallel to the anisotropy direction $z$, is defined by the Hamiltonian,
\begin{equation}
\label{hamh}
  H_\mathrm{XXZ}=\sum_{m=1}^M \Big\{ \sigma^x_m \sigma^x_{m+1} +
  \sigma^y_m\sigma^y_{m+1} + \Delta(\sigma^z_m\sigma^z_{m+1}-1) - {h \over 2} \sigma^z_m\Big\},
\end{equation}
where $\s^a_m,\ a=x,y,z,$ are the Pauli matrices associated to the site 
$m$ of the chain and acting in the two-dimensional space $\mathcal{H}_m$ at 
site $m$, $\Delta$ is the anisotropy parameter and $h$ is a constant magnetic field in the $z$ direction.

The eigenstates and the spectrum of this model have been described using the Bethe ansatz method in \cite{Orb58,YanY66a}, and later by its algebraic version in \cite{FadST79}.

The aim of this article is to describe the $n$-point correlation functions of the local spin operators of this model in the thermodynamic limit, in the presence of a constant magnetic field $h$, for the anisotropy parameter $\Delta \ge -1$ and at zero temperature. If the anisotropy parameter is less than $-1$, the ground state of the Hamiltonian (\ref{hamh})  is ferromagnetic and all such correlation functions  can be calculated easily. In fact let us denote by $\ket{\psi_g}$ the ground state in the massless regime and any one of two
ground states constructed by the algebraic Bethe ansatz in the massive regime. 
Let $E^{\e'_m,\e_m}_{m}$ be the elementary operators acting on $\mathcal{H}_m$ at site $m$ as the $2 \times 2$ matrices   $E^{\e',\e}_{lk}=\delta_{l,\e'}\delta_{k,\e}$, and $\pl_{j=1}^m E^{\e'_j,\e_j}_j$ any product of such elementary operators from site one to 
$m$. We will compute the following correlation functions,
\begin{equation}
F_m(\{\e_j,\e'_j\})=\frac {\bra{\psi_g}\pl_{j=1}^m 
E^{\e'_j,\e_j}_j\ket{\psi_g}}{\bra{\psi_g}\psi_g\r} .
\label{genabcd00}
\end{equation}
Any $n$-point correlation function can be obtained from these building blocks.

Our approach is based on the algebraic Bethe ansatz formulation of this model \cite{FadST79} and the actual resolution of its associated quantum inverse scattering problem for local spins given in \cite{KitMT99}. The first results in the thermodynamic limit were obtained using this method for the spontaneous magnetization of the XXZ model in \cite{IzeKMT99}.

Previous results about correlation functions were obtained in \cite{JimM95L} and \cite{EssFIK95}. In the case $h = 0$ our results agree with the one following the q-deformed Knizhnik-Zamolodchikov (KZ) equations \cite{JimM95L} in both the massless ($-1<\Delta<1$) \cite{JimM96} and massive ($ \Delta \ge 1$) \cite{JimMMN92} regimes.

Before describing our method, let us first briefly recall the problem to be solved, the main difficulties to overcome, and the approaches that have been used previously in this context.

In fact, the computation of exact correlation functions is one of the most challenging and longstanding problem in the domain of integrable models of field theory and statistical mechanics. This is a central question both from the fundamental and mathematical  point of view, and also when one is willing to apply such models for example to  condensed matter phenomena. 

Despite the great amount of remarkable works (see \cite{BogIK93L,JimM95L,Mcc68} and references therein), no general method for computing exact and explicit  manageable expressions for the correlation functions of generic integrable models is available at present. 

In this context, the  XXZ Heisenberg spin-$1 \over 2$ chain in the anti-ferromagnetic regime has already been used as a representative example of integrable model, with non-trivial interaction and bound states (in particular it is not equivalent to free fermions, except at $\Delta = 0$), associated to a q-deformed trigonometric $R$-matrix, and which contains all the basic difficulties in computing correlation functions. 

In fact, while the ground state of such models in the thermodynamic limit can be described very precisely, for example  by means of the  Bethe ansatz method, the main difficulty in the computation of the correlation functions lies precisely in the highly complicated structure of such a state and its associated excited states. Indeed, for generic, really interacting models of this class like the XXZ Heisenberg spin-$1 \over 2$ chain in the regime $\Delta \ge -1$, the ground state is given as an extremely non-local action of the local fields (the spins on each lattice site here) of the model on a known reference state (here the ferromagnetic state with all spin up). 

This fact can be seen easily, by using the algebraic Bethe ansatz approach \cite{FadST79}. The main object of this method is the so-called quantum monodromy matrix. Its matrix elements are operators acting in the space of states of the model, its trace (the transfer matrix) containing a set of conserved charges including the Hamiltonian, while the other elements lead to the creation and annihilation operators of the Bethe eigenstates. The quantum monodromy matrix is constructed as an ordered matrix product all along the 
chain of the so-called quantum Lax operators defined in terms of the local  
spin operators at each site of the lattice. Hence, the monodromy matrix 
itself, and the creation and 
annihilation operators of the Bethe eigenstates, are highly non-local expressions in terms of these local spin operators. In particular, let us mention that such operators are  sums of up to $2^{M}$ terms, each being the product of $M$ local spin operators all along the chain of $M$ sites. Then, 
the ground state in the regime $\Delta \ge -1$ is obtained by acting $N$ 
times with one of the monodromy matrix element ($M = 2N$ for example), its 
local description being henceforth a very complicated combinatorial problem. 

Hence, although on the one hand the local spin operators satisfy a simple algebra, and on the other hand, the elements of the quantum monodromy matrix obey the well-known quadratic (Yang-Baxter) relations, the interplay between these two sets of operators is a priori quite involved. In particular, the commutation relations between the local spin operators and the quantum 
monodromy matrix elements do not have a simple and manageable form. This 
fact, also related to the non-local structure of the ground state (and in 
fact of all the other eigenstates), makes the computation of correlation functions very difficult.

Indeed, until very recently, only very few models were known for which correlation function can be computed in an exact and explicit way. The typical examples are the Ising model, related to free fermions, and conformal field theories dealing with critical or massless systems in the continuum. 

Beyond these models, in the framework of integrable systems solvable by means 
of Bethe ansatz 
\cite{Len64,Len66,FadST79,Fad82,Bax82L,Gau83L,BogIK93L},  related to a quantum group structure \cite{Dri87,Jim85,Jim86,FadRT90} and associated to an R-matrix solving the Yang-Baxter equation, two main approaches have been designed to deal with this problem.

One of them, described in the book 
\cite{BogIK93L}, is based on  the algebraic Bethe ansatz method \cite{FadST79,Fad82,Bax82L,BogIK93L} applied to models defined in a finite volume (or finite lattice). Determinant representations of correlation functions are obtained \cite{IzeK84,IzeK85,EssFIK95,BogIK93L},  containing however vacuum expectation values of auxiliary (quantum) ``dual fields'', which cannot be eliminated in general from the final result. Hence, explicit expressions for the correlators in the thermodynamic limit cannot be obtained directly from this approach. Instead, the strategy is to embed these determinant formulas in systems of 
integrable integro-difference equations from which only large distance 
asymptotics of some correlation functions can be extracted from the resolution of (matrix) operator valued Riemann-Hilbert 
problems. Note however, that for spin chains no such asymptotics have been determined yet, except for the free fermion points \cite{ItsIKS93}. 

The other approach relies on the study of  form factors and correlation functions of quantum integrable models 
directly in the infinite volume (or infinite lattice) limit. For quantum field theory models, it started with the 
study of analytic properties and bootstrap equations  for the factorized S-matrices and form factors of integrable 
quantum field theories in infinite volume \cite{ZamZ79,Smi92L,KarW78}. A typical model here is the two-dimensional 
Sine-Gordon relativistic quantum field theory. There it was realized that the set of equations satisfied 
by the form factors are closely related to the q-deformed Knizhnik-Zamolodchikov equations arising 
from representation theory of quantum affine algebras, and their q-deformed vertex operators 
\cite{FreR92,DatJO93,JimKMQ94,Smi92,Smi93,Smi93a,Var94}. For lattice models, such as the the
 XXZ Heisenberg spin-$1 \over 2$ (infinite) chain, it uses the corner transfer matrix introduced 
by Baxter \cite{Bax76,Bax77,Bax82L,Tha86} in the context of integrable models of statistical mechanics,
 and  very plausible hypothesis about the representation of the Hamiltonian of the model  as a central
 element of the corresponding quantum affine algebra (here ${\cal U}_q ({\hat {sl_2}})$) in the infinite 
lattice, the space of states being identified in terms of its
 highest weight  modules \cite{JimMMN92,JimMN93,JimM95L}. Form factors and correlation 
functions are then described in terms of q-deformed vertex operators, 
leading via bosonization \cite{KatQS93}, to multiple integral formulas for them.
 
In \cite{KitMT99} we started the development of a new approach to this problem in the example of the XXZ Heisenberg
 spin-$1 \over 2$ finite chain. 
In the present article we use our results \cite{KitMT99} for the finite chain to obtain, in the thermodynamic limit, 
multiple integral representations of arbitrary $n$-points correlation functions of the XXZ Heisenberg
 spin-$1 \over 2$ in a constant magnetic field. In the zero magnetic field case, our results agree with the one
 described in \cite{JimM95L} for the massive regime, and with \cite{JimM96} for the massless regime. However,
 being  grounded on the algebraic Bethe ansatz method, we are able to deal with the case of a constant magnetic
 field, a situation where the infinite quantum affine symmetry algebra used in \cite{JimM95L} breaks down.

Although developed in the context of the XXZ Heisenberg spin-$1 \over 2$ chain, we expect that the general strategy
 we used could be applicable to other 
models. Our approach to the computation of correlation functions starts with the above general observation that the
 main difficulty to be solved is the fact that two types of operators are mixed in such quantities : the local 
(spin) operators and the (highly) non-local (in terms of the local spins) creation/annihilation operators of the 
eigenstates of the Hamiltonian. The solution to this problem is to express both types of operators, 
in a simple and manageable way, into a common algebraic structure in terms of which the corresponding correlation
 functions can be evaluated. It is this strategy that has been already used in fact in all cases where explicit
 results have been obtained for correlation functions : for example in  the Ising model using Clifford algebra 
\cite{SatMJ80} or for the XXZ Heisenberg spin-$1 \over 2$ (infinite) chain in the massive regime and in a zero magnetic
 field by using the ${\cal U}_q ({\hat {sl_2}})$ algebra and its associated $q$-deformed vertex 
 operators \cite{JimM95L}.
Here we will solve this problem first for the finite XXZ Heisenberg spin-$1 \over 2$ chain by using the Yang-Baxter
 algebra generated by the quantum monodromy $2 \times 2$ matrix, 
\[T(\la)
        =\left(\ba{cc}
                    A(\la)& B(\la)\\ 
                    C(\la)& D(\la)\ea\right).\]
The ground state of the model can be constructed by means of algebraic Bethe ansatz by the successive actions of the operators $B(\la_k)$ on the ferromagnetic reference state with all spin up, for a particular set of spectral parameters $\{\la_k\}$ solving the Bethe equations. 

 Our method is based on the algebraic Bethe ansatz and  goes along the following main steps. 
The ground state $\bra{\psi_g}$ of the XXZ Heisenberg spin-$1 \over 2$ finite chain is given 
as the successive  action of the operators $C(\la_k)$ (resp. $B(\la_k)$) (elements of the
 quantum monodromy matrix) on the ferromagnetic reference state $\bra{0}$ (resp. $\ket{0}$),
 the state with all spin up,  for a particular set of spectral parameters $\{\la_k\}$ solving
 the Bethe equations, namely, $\bra{\psi_g} = \bra{0} \pl_k C(\la_k) $ and  
$\ket{\psi_g} = \pl_k B(\la_k) \ket{0}$.
Our main point is that  any local elementary operator $E^{\e'_m,\e_m}_{m}$ can be expressed in terms of the operators entries of the above quantum monodromy matrix \cite{KitMT99} (namely by actually solving the quantum inverse scattering problem) as,
\[E^{\e'_j,\e_j}_j=\pl_{k=1}^{j-1}\Bigl(A(\xi_k)+D(\xi_k)\Bigr)
T_{\e_j,\e'_j}(\xi_j)\pl_{k=j+1}^{M}\Bigl(A(\xi_k)+D(\xi_k)\Bigr).\]
 Then, the elementary building blocks of the correlation functions, 
\begin{equation}\label{problem0}
    \bra{\psi_g}  \biggl( \pl_i E_i^{\e'_i \e_i} \biggr)
                    \ket{\psi_g}
\end{equation} 
can  be evaluated as the scalar product of the ground state $\ket{\psi_g}$
 with the state $\bra{\psi_g}  
\biggl( \pl_i E_i^{\e'_i \e_i} \biggr)$.
To compute such a quantity, we  first evaluate the state $\bra{\psi_g} \biggl( \pl_i E_i^{\e'_i \e_i} \biggr)$. There,  $\biggl( \pl_i E_i^{\e'_i \e_i} \biggr)$ is replaced by a certain products of elements of the quantum monodromy matrix. Using their known (Yang-Baxter) commutation relations with the operators  $C(\la_k)$, it is possible to obtain the above wanted state as the following linear combination of states,
\begin{equation}\label{1re}
 \bra{\psi_g}  \biggl( \pl_i E_i^{\e'_i \e_i} \biggr)
                     = \sul_{i \in I} \alpha_i \bra{0} \pl_{k \in K_i} C(\la_k),
\end{equation} 
with some (computable) coefficients $\alpha_i$. It then remains to evaluate the scalar product of the ground state $\ket{\psi_g}$ with any state in this sum. This problem was also solved in a compact form in \cite{KitMT99} 
(see also \cite{Sla89}) as the ratio of two explicit determinants. The next step is to compute the resulting sums in the thermodynamic limit. There, we use these determinant representations of the scalar products, to finally obtain multiple integral formulas, containing as integration measure, the determinant of  the derivatives of the spectral density $\rho$ characterizing the ground state, and solution of the Lieb equation. As a consequence, we can obtain the result also in the presence of a constant magnetic field.

The article is organized as follows.
After briefly describing the algebraic Bethe ansatz approach to the XXZ Heisenberg 
spin-$1 \over 2$ model in section 2, the general strategy and tools  of our method 
are explained in section 3. In section 4, we apply this scheme to the so-called 
"emptiness formation probability", which is the probability to find a ferromagnetic 
configuration of length $m$ in the massless or massive anti-ferromagnetic ground state. 
The computation is given first for a zero magnetic field in order to explain the basic 
techniques of the thermodynamic limit. Then in the next section, we derive the multiple 
integral formulas for an arbitrary $n$ point correlation function, also for the zero 
magnetic field case. We generalize this result in section 6, to the constant magnetic 
field situation.


\section{The XXZ spin-$\frac{1}{2}$ Heisenberg chain}

The Hamiltonian of the chain of finite length $M$~\cite{Hei28,Bet31} is given 
by,
\begin{equation}
\label{ham}
  H_\mathrm{XXZ}=\sum_{m=1}^M \Big\{ \sigma^x_m \sigma^x_{m+1} +
  \sigma^y_m\sigma^y_{m+1} + \Delta(\sigma^z_m\sigma^z_{m+1}-1)\Big\},
\end{equation}
and we impose periodic boundary conditions.
Here $\s^a_m,\ a=x,y,z,$ are the Pauli spin operators
acting in the local quantum spin $\frac{1}{2}$ space $\mathcal{H}_m$ 
at site $m$.
The anisotropy parameter $\Delta$ defines the physical nature of the model:
when $\Delta\le -1$, the ground state of the Hamiltonian is ferromagnetic,
whereas its magnetization is equal to zero when $\Delta>-1$. We shall focus
our attention on this last domain, which itself decomposes into a massive
regime (for $\Delta>1$), and a gapless regime (for $-1<\Delta<1$) in the 
thermodynamic limit ($M\rightarrow\infty$).

The  $R$-matrix of the XXZ model is,
\begin{equation}
   R(\la,\mu)=
  \left(\ba{cccc}1&0&0&0\\
                 0&b(\la,\mu)&c(\la,\mu)&0\\
                 0&c(\la,\mu)&b(\la,\mu)&0\\
                 0&0&0&1\ea\right),
\end{equation}
where the functions $b(\la,\mu)$ and $c(\la,\mu)$ are defined as,
\[b(\la,\mu)=\frac{\sinh(\la-\mu)}{\sinh(\la-\mu+\eta)},
\quad c(\la,\mu)=\frac{\sinh\eta}{\sinh(\la-\mu+\eta)}.\] 
The parameter $\eta$ is here related to the anisotropy parameter 
$\Delta$ of the Hamiltonian by,
\begin{equation*}
  \Delta=\frac{1}{2}(q+q^{-1}),\quad \text{with}\ q=e^{\eta}.
\end{equation*}
The $R$-matrix is a linear operator in the tensor product of two two-dimensional
linear spaces $V_1 \otimes V_2$, where each $V_i$ is isomorphic to 
${\mathbb C}^2$, and depends generically on two spectral parameters $\lambda_1$ and $\lambda_2$
 associated to these two vector spaces. It is denoted by $R_{12} (\lambda_1, \lambda_2)$. 
Such an $R$-matrix satisfies the Yang-Baxter equation.

Identifying one of the two vector spaces of the $R$-matrix with the quantum
space $\mathcal{H}_m$, one defines the quantum $L$-operator of the inhomogeneous
chain at site $m$ as,
\begin{equation}
   L_m(\lambda, \xi_m) = R_{0m} (\lambda -\xi_m),
\end{equation}
where $\xi_m$ is an arbitrary inhomogeneity parameter attached to the site $m$.
Here $R_{0m}$ acts in $V_0 \otimes \mathcal{H}_m$, where $V_0$ is
an auxiliary space isomorphic to $\mathbb{C}^2$.
The monodromy matrix is constructed as an ordered product of such 
$L$-operators:
\[T(\la)=R_{0 M}(\la-\xi_M)\dots 
           R_{0 2}(\la-\xi_2)R_{0 1}(\la-\xi_1)
        =\left(\ba{cc}
                    A(\la)& B(\la)\\ 
                    C(\la)& D(\la)\ea\right)_{[0]}.\]
In the last formula, the monodromy matrix is represented as a $2\times 2$ matrix
in the auxiliary space $V_0$, whose entries 
$A(\la)$, $B(\la)$, $C(\la)$, and $D(\la)$ are operators
in the quantum space $\mathcal{H}$ of the chain. 

The transfer matrix 
$\mathcal{T}(\la)$ is defined as the trace $A(\la)+D(\la)$ of the monodromy matrix.
Transfer matrices commute with each other for different values of the spectral
parameter $\la$. They commute  also 
with the Hamiltonian (\ref{ham}) in the homogeneous case where all
$\xi_m$ are equal  as the Hamiltonian can be reconstructed in terms of the transfer 
matrix by means of the following ``trace identity'':
\begin{equation}
H_{\mathrm{XXZ}}=2\sinh\eta\left.\pd{\la}\log\mathcal{T}(\la)\right|_{\la=\xi_j}+\mathrm{const}.
\end{equation}

Common eigenstates of the transfer matrices (and thus of the 
Hamiltonian~\eqref{ham} in the homogeneous case) 
can be constructed by successive actions
of operators $B(\la)$ on the reference state $\ket{0}$, which is the 
ferromagnetic state 
with all the spins up. More precisely, the state 
$B(\la_1)\dots B(\la_N)\ket{0}$
is a common eigenstate of the transfer matrices if the set of spectral
parameters $\{\la_j\}_{1 \le j \le N}$ is a solution of the Bethe equations,
\begin{equation}
   \frac{a(\lambda_j)}{d(\lambda_j)}\prod\begin{Sb}k=1\\k \not= j\end{Sb}^N
      \frac{b(\lambda_j, \lambda_k)}{b(\lambda_k, \lambda_j)} 
   = 1,                    \qquad 1 \le j \le N, 
\label{eq:bethe}
\end{equation}
where $a(\la)=1$ and $d(\la)=\prod_{i=1}^M b(\la, \xi_i)$ 
are the eigenvalues of operators $A(\la)$ and 
$D(\la)$ respectively on the reference state $\ket{0}$.
The corresponding eigenvalue for the transfer matrix $\mathcal{T} (\mu)$ is then,
\begin{equation}\label{theta}
   \tau(\mu, \{\lambda_j\})=
      a(\mu)\prod_{j=1}^n b^{-1}(\lambda_j, \mu) 
      + d(\mu)\prod_{j=1}^n b^{-1}(\mu, \lambda_j). 
\end{equation}

The Bethe equations can also be written in a logarithmic form:
\begin{equation}
   M p_{0_{\text{tot}}}(\la_j) + \sul_{k=1}^N\theta(\la_j-\la_k)=2\pi n_j,
     \quad 1\le j \le N, 
\label{bethelog}
\end{equation}
where $n_j$ are integers for $N$ odd and half integers for $N$ even. 
The bare momentum $p_{0_{\text{tot}}}(\la)$ and the scattering phase $\theta(\la)$ are defined as,
\begin{align*}
  p_{0_{\text{tot}}}(\la) &=\frac{i}{M}\ln\frac{d(\la)}{a(\la)}
           =\frac{1}{M} \sul_{k=1}^M p_0 (\la-\xi_k+\frac{\eta}{2}),\\
  p_0(\la) &=i\ln\frac{\sinh(\la-\frac{\eta}{2})}{\sinh(\la+\frac{\eta}{2})},\\ 
  \theta(\la)&=i\ln\frac{\sinh(\eta+\la)}{\sinh(\eta-\la)}.
\end{align*}
In the thermodynamic limit ($M\rightarrow\infty$), these Bethe equations for
the ground state become
an integral equation for the quasi-particle density $\rho$ in the rapidity representation
(Lieb equation)~\cite{LieL63,YanY66a}:
\begin{equation}
   \rho_{\text{tot}}(\alpha)+\int_{-\Lambda}^{\Lambda} K(\alpha-\beta)\rho_{\text{tot}}(\beta)
\, d\beta
               =\frac{p_{0_{\text{tot}}}'(\alpha)}{2\pi},
  \label{Lieb}
\end{equation}
where the new real variables $\alpha$ are defined in terms of general spectral parameters $\lambda$
differently in the two domains:
\begin{align*}
   &\alpha=\lambda \quad\ \text{for}\quad -1<\Delta<1,\\
   &\alpha=i\lambda \quad \text{for}\quad\ \Delta>1.    
\end{align*}
The density $\rho$ is defined as the limit of the quantity 
$\frac{1}{M(\alpha_{j+1}-\alpha_j)}$, and the functions $K(\alpha)$ and 
$p_{0_{\text{tot}}}'(\alpha)$ are the
derivatives with respect to $\alpha$ of the functions 
$-\frac{\theta(\lambda(\alpha))}{2\pi}$ and $p_{0_{\text{tot}}}(\lambda(\alpha))$:
\begin{align}
 & \begin{aligned}
    K(\alpha) = &\frac{\sin 2\zeta}
                     {2\pi \, \sinh (\alpha +i\zeta) \sinh(\alpha -i\zeta)}\\
    p_0'(\alpha) = &\frac{\sin \zeta}
              {\sinh(\alpha+i\frac{\zeta}{2}) \sinh(\alpha-i\frac{\zeta}{2})}
  \end{aligned}
   \qquad  \text{for}\ -1<\Delta<1,\ \text{with}\ \zeta = i\eta,\\
  & \begin{aligned}
    K(\alpha) = &\frac{\sinh 2\zeta}
                     {2\pi \, \sin (\alpha +i\zeta) \sin(\alpha -i\zeta)}\\
    p_0'(\alpha) = &\frac{\sinh \zeta}
              {\sin(\alpha+i\frac{\zeta}{2}) \sin(\alpha-i\frac{\zeta}{2})}
  \end{aligned}
   \qquad \quad \text{for}\ \ \ \Delta>1,\ \text{with}\ \zeta = -\eta,\\
  &\text{with}\quad p_{0_{\text{tot}}}'(\alpha) = \frac{1}{M} \sul_{i=1}^M 
                  p_0'(\alpha-\beta_k-i\frac{\zeta}{2}),
\end{align}
where $\beta_k=\xi_k$ in the domain $-1<\Delta<1$, and $\beta_k=i\xi_k$
in the domain $\Delta>1$.
 The 
integration limit $\Lambda$ is equal to $\frac{\pi}{2}$ for $\Delta>1$, 
and to $+\infty$ for $-1<\Delta<1$.

The solution for the Lieb equation~\eqref{Lieb} in the homogeneous model where
all parameters $\xi_k$ are equal to $\eta / 2$, that is the density for the
ground state of the Hamiltonian~\eqref{ham} in the thermodynamic limit, is
given by the following function~\cite{YanY66a}:
\begin{alignat}{2}
   \rho(\alpha) &= \frac{1}{2\zeta \cosh (\frac{\pi\alpha}{\zeta})}\quad &
       \text{for}&\ -1<\Delta<1,
   \label{dens1}    \\
   \rho(\alpha) &= \frac{1}{2\pi} \sul_{n=-\infty}^{+\infty} 
                   \frac{e^{2in\alpha}}{\cosh(n\zeta)}\quad &
       \text{for}&\quad \Delta>1.
\label{dens2}
\end{alignat}
For technical convenience, we will also use in the following the solution
of the inhomogeneous Lieb equation, that is the function,
\begin{equation}
  \rho_{\text{tot}}(\alpha) =\frac{1}{M} \sul_{i=1}^M 
                  \rho(\alpha-\beta_k-i\frac{\zeta}{2}).
\end{equation}
It will be also convenient to consider, without
any loss of generality, that the inhomogeneity parameters are contained
in the region $-\zeta<\mathrm{Im} \beta_j < 0$.

Let us mention at last that
the ground state of the XXZ model in the region $\Delta>1$
 is degenerated in the thermodynamic limit ($M\rightarrow\infty$), 
namely there are two 
states with the same energy 
(and characterized by the same density~\eqref{dens2}),  which we will call the 
ground state $\ket{\Psi_1}$ and the quasi-ground state 
$\ket{\Psi_2}$ (on the finite lattice, these states possess 
different energy). In this domain, the correlation function at zero temperature
is thus half of the trace on these two states, that is of the sum of the two 
corresponding matrix elements.
In the domain $-1<\Delta<1$, the ground state is not degenerated.


\section{Quantum inverse scattering problem and correlation functions}

In this section, we explain in more detail our general procedure to compute correlation 
functions
of the XXZ chain in the algebraic Bethe ansatz framework, along the lines described in the 
introduction. 
Our method is based on
the study of the finite chain which has been performed in~\cite{KitMT99}.
We recall here the main
results of~\cite{KitMT99}.

Our purpose is to compute, in the algebraic Bethe ansatz formalism, general
matrix elements of products of local spin operators,
\begin{equation}\label{problem}
    \bra{0} \pl_j C(\mu_j) \biggl( \pl_i \sigma_i^{\eps_i} \biggr)
                   \pl_k B(\la_k) \ket{0}
\end{equation} 
between two Bethe states $\bra{0} \pl_j C(\mu_j)$ and $\pl_k B(\la_k) \ket{0}$.
Here $\sigma_i^{\eps_i}$, $\eps_i \in \{+,-,z\}$, are Pauli spin operators 
at site $i$.
The different difficulties which appear here (and which we  described in  the introduction) lead us to decompose the computation of such correlation functions into four main steps.

\subsection{Solution of the quantum inverse problem}

The algebraic Bethe ansatz method is based on the  
commutation relations
given by the $R$-matrix for the generators $A$, $B$, $C$, $D$ of the 
Yang-Baxter algebra. The first combinatorial problem which appears in the
study of expressions of the form~\eqref{problem} is that they contain both
local operators (spin operators) and highly non-local ones (operators $B$ and
$C$) and the relations between these two types of operators
are not clear. One needs to link these two types of operators. One way to do
it, in the spirit of the classical inverse scattering method, 
is to solve the quantum inverse scattering problem for the chain, that is to
express the local quantum spin operators in terms of the generators $A$, $B$,
$C$, $D$ of the Yang-Baxter algebra. This has been done in~\cite{KitMT99},
and we recall here the result.

\begin{theorem}\label{thm:reconstr}
    Local spin operators at a given site $i$ of the inhomogeneous XXX or XXZ
   Heisenberg chain are given by,
   \begin{align}
     \sgm_i &= \prod_{\alpha=1}^{i-1} \left( A + D \right) (\xi_{\alpha}) 
               \ \cdot\
               B(\xi_i) \ \cdot\
               \prod_{\alpha=i+1}^N \left( A + D \right) (\xi_{\alpha}),\label{s-B}\\
     \sgp_i &= \prod_{\alpha=1}^{i-1} \left( A + D \right) (\xi_{\alpha}) 
               \ \cdot\
               C(\xi_i) \ \cdot\
               \prod_{\alpha=i+1}^N \left( A + D \right) (\xi_{\alpha}),\label{s+C}\\
     \sgz_i &= \prod_{\alpha=1}^{i-1} \left( A + D \right) (\xi_{\alpha}) 
               \ \cdot\
               (A-D) (\xi_i) \ \cdot\
               \prod_{\alpha=i+1}^N \left( A + D \right) (\xi_{\alpha}). \label{szA-D}      
   \end{align}
\end{theorem}

This reduces our problem 
to the mere computation of matrix elements of 
products of $A$, $B$, $C$, $D$ operators in the reference state, for which
commutation relations of the Yang-Baxter algebra can be used.
 
Indeed, for any integer
$k$ and any subset $\{ i_j \}_{1\le j \le k}$ of $\{ 1,\ldots, N\}$,
with the convention $i_1< i_2 \ldots < i_k$, the correlation function
for spins at sites $i_1,\ldots, i_k$ between two Bethe states
$\bra{0}\ C(\mu_1) \ldots C(\mu_{n_1})$ and 
$B(\lambda_1)\ldots B(\lambda_{n_2})\ \ket{0}$
has the following
form:
\begin{multline}\label{corrABCD}
    \bra{0}\ C(\mu_1) \ldots C(\mu_{n_1})\ \sg^{\eps_1}_{i_1}\ 
               \sg^{\eps_2}_{i_2}\/
               \ldots\/\sg^{\eps_k}_{i_k}\
     B(\lambda_1)\ldots B(\lambda_{n_2})\ \ket{0}=\\
   = \prod_{\alpha=1}^{i_1-1} \prod_{j=1}^{n_1} b^{-1}(\mu_j, \xi_\alpha)\ \cdot
     \prod_{\alpha=i_k+1}^{N} \prod_{j=1}^{n_2} b^{-1}(\lambda_j, \xi_\alpha)\ 
       \times\\ 
   \times\ \bra{0}\ C(\mu_1) \ldots C(\mu_{n_1})\ \cdot\ 
          X^{\eps_1} (\xi_{i_1})\ \cdot
     \prod_{\alpha=i_1+1}^{i_2-1} \bigl( A+D \bigr) (\xi_\alpha)\ \cdot\
                   X^{\eps_2} (\xi_{i_2})\ \ldots\\
    \ldots  \prod_{\alpha=i_{k-1}+1}^{i_k-1} \bigl( A+D \bigr) (\xi_\alpha)\ 
                       \cdot\
                            X^{\eps_k} (\xi_{i_k})\ \cdot\
      B(\lambda_1)\ldots B(\lambda_{n_2})\ \ket{0},
\end{multline}
where $\eps_j,\ 1 \le j \le k$, takes the values $+,\ -$, or $z$,
$X^{\eps_j}$ being equal respectively to $C,\ B$ and $A-D$.

\subsection{Action of operators $A$, $B$, $C$, $D$ on a general state}

The second step is thus to express the successive action of any product
of  $A$, $B$, $C$, $D$ operators on a state constructed by action of 
$C$ operators on the reference states. Action of $A$, $B$, $C$, $D$
on such a state are well known (see for example~\cite{BogIK93L}), but
we recall them here in a more convenient form for our purpose. 

The action of the operators $A(\la)$ and  $D(\la)$ 
on the states constructed by successive actions of  operators $C(\la)$ can be 
written in the following form: 
\begin{align}
 \bra{0}\pl_{k=1}^{N}C(\la_k)\,A(\la_{N+1})=&
  \sul_{a'=1}^{N+1} a(\la_{a'}) 
   \frac{\pl_{k=1}^N   
   \sinh(\la_k-\la_{a'}+\eta)}{\pl_{k=1\atop{k\neq a'}}^{N+1}\sinh(\la_k-\la_{a'})}
   \,\bra{0}\pl_{k=1\atop{k\neq a'}}^{N+1} C(\la_k);\label{abbb}\\
\bra{0}\pl_{k=1}^{N}C(\la_k)\, D(\la_{N+1})=&
  \sul_{a=1}^{N+1} d(\la_a) 
   \frac{\pl_{k=1}^N   
   \sinh(\la_a-\la_k+\eta)}{\pl_{k=1\atop{k\neq a}}^{N+1}\sinh(\la_a-\la_k)}
   \,\bra{0}\pl_{k=1\atop{k\neq a}}^{N+1} C(\la_k)
   .\label{dbbb}\end{align}
The action of the operator $B(\la)$ is more complicated:
\begin{align}
\bra{0}\pl_{k=1}^N C(\la_k)\, B(\la_{N+1})=&\sul_{a=1}^{N+1} 
  d(\la_a)\frac{\pl_{k=1}^N   
   \sinh(\la_a-\la_k+\eta)}{\pl_{k=1\atop{k\neq a}}^{N+1}\sinh(\la_a-\la_k)}
\times\nonumber\\
\times&\sul_{a'=1\atop{a'\neq a}}^{N+1} \frac{a(\la_{a'})}{\sinh(\la_{N+1}-\la_{a'}+\eta)}
\frac{\pl_{j=1\atop{j\neq a}}^{N+1}   
   \sinh(\la_j-\la_{a'}+\eta)}{\pl_{j=1\atop{j\neq a,a'}}^{N+1}\sinh(\la_j-\la_{a'})}
 \bra{0} \pl_{k=1\atop{k\neq a,a'}}^{N+1} C(\la_k), \label{cbbb}
\end{align} 
but in the case which is
interesting for the computation of the correlation functions when $\la_{N+1}=\xi_k$ and
hence $d(\la_{N+1})=0$, we obtain a more simple result:
       \begin{align}
\bra{0}\pl_{k=1}^N C(\la_k) B(\la_{N+1})=&\sul_{a=1}^{N} 
  d(\la_a)\frac{\pl_{k=1}^N   
   \sinh(\la_a-\la_k+\eta)}{\pl_{k=1\atop{k\neq a}}^{N+1}\sinh(\la_a-\la_k)}
\times\nonumber\\
\times&\sul_{a'=1\atop{a'\neq a}}^{N+1} a(\la_{a'})\frac{\pl_{j=1\atop{j\neq a}}^N   
   \sinh(\la_j-\la_{a'}+\eta)}{\pl_{j=1\atop{j\neq a,a'}}^{N+1}\sinh(\la_j-\la_{a'})}
 \bra{0} \pl_{k=1\atop{k\neq a,a'}}^{N+1} C(\la_k). \label{cbbb1}
\end{align}
It should be mentioned that the action of $B$ is similar to the successive action of $D$ and $A$. Using these formulae, one can reduce expressions of the form~\eqref{corrABCD}
to sums of scalar products of a Bethe state with an arbitrary state constructed
by successive actions of $B$ operators on the reference state.

\subsection{Scalar products}

The third step of the computation is thus to find an explicit and 
convenient expression for such scalar products. Usual Bethe ansatz techniques,
based only on the use of commutation relations, generally generate huge sums
which are difficult to sum up. In~\cite{KitMT99}, a direct computation in
a new basis ($F$-basis)~\cite{MaiS96} has been performed and leads 
to an explicit expression for such scalar
products as a determinant of usual functions of the model:

\begin{theorem}
  Let  $\{\la_1,\dots,\la_N\}$ be a solution 
  of the Bethe equations~\eqref{eq:bethe} and $\{\mu_1,\dots,\mu_N\}$
  be an  arbitrary set of parameters. Then the scalar product, 
   \begin{equation}
      S_N(\{\mu_j\},\{\la_k\}) = 
  \bra{0}\ \prod_{j=1}^N C (\mu_j) \ 
        \prod_{k=1}^N B (\lambda_k)\ \ket{0}
   \end{equation} 
  can be represented as a ratio
  of two determinants,
  \begin{equation}
    S_N(\{\mu_j\},\{\la_k\})=S_N(\{\la_k\},\{\mu_j\})
    =\frac{\mathrm{det} T(\{\mu_j\},\{\la_k\})}
          {\mathrm{det} V(\{\mu_j\},\{\la_k\})},
    \label{theorem1}
  \end{equation}
  of the following $N \times N$ matrices $T$ and $V$:  
  \begin{equation}
    T_{ab}=\pd{\la_a}\tau(\mu_b,\{\la_k\}),
    \qquad V_{ab}=\frac 1{\sinh(\mu_b-\la_a)},\qquad 1\le a,b \le N,
    \label{tv}
  \end{equation}
where $\tau(\mu_b,\{\la_k\})$ is the eigenvalue of the transfer matrix $\mathcal{T}(\mu_b)$
corresponding to  the Bethe state $\prod_{k=1}^N B (\lambda_k)\ \ket{0}$ given by (\ref{theta}).
  \label{theor} 
\end{theorem}

This result is equivalent to   the scalar product  formula obtained in \cite{Sla89}.

When particularizing  this formula in the case when the two states are
equal, one obtains the Gaudin formula for the norm of a Bethe state:

\begin{equation}
   \bra{0}\ \prod_{j=1}^N C (\la_j) \ 
           \prod_{k=1}^N B (\la_k)\ \ket{0}
        = \sinh^N \eta \prod_{\alpha\ne\beta} 
           \frac{\sinh(\la_\alpha- \la_\beta+\eta)}
                {\sinh(\la_\alpha- \la_\beta)}
            \det \Phi'(\{\la_\a\}),
\label{gaudin}
\end{equation}
where $\Phi'$ is a $N \times N$ matrix the elements of which are given by:
\begin{align}
  \Phi_{ab}' &= -\partialsur\la_b 
                       \ln\biggl(\frac{a(\la_a)}{d(\la_a)}
                       \prod_{k=1 \atop k\ne a}^N
                       \frac{b(\la_a, \la_k)}{b(\la_k, \la_a)}
                       \biggr).
\label{matrix-gaudin}
\end{align}

By means of this expression for the scalar product, 
general correlation functions for the finite chain can now be expressed 
as sums of determinants. 

\subsection{Thermodynamic limit}

The last step of our method, to obtain the general correlation functions in the
infinite volume limit, is to take the thermodynamic limit of the expressions
obtained for the finite chain. This has already been done in a particular
case in the article~\cite{IzeKMT99}, where the Baxter formula 
for the spontaneous staggered magnetization in the domain $\Delta>1$ has
been derived
by this method. Here, we generalize it to any $n$-point correlation function.

In the thermodynamic limit $M\rightarrow \infty$, the 
Bethe equations for the
ground state become the integral Lieb equation~\eqref{Lieb} for the density.
In a more general way, 
for any $\mathcal{C}^\infty$ function $f$ ($\pi$-periodic
in the domain $\Delta>1$), sums over all the values of $f$ at the point 
$\alpha_j$, $1\le j \le N$, parameterizing the ground state, can be replaced
in the thermodynamic limit by an integral involving the density 
$\rho$ solution of the Lieb equation~\cite{IzeKMT99}:
\begin{equation}
   \frac{1}{M} \sul_{j=1}^N f(\alpha_j)= 
    \int_{-\Lambda}^{\Lambda} f(\alpha) \rho_{\text{tot}}(\alpha)\, d\alpha 
    + O(M^{-\infty}).
\label{sumint}
\end{equation} 
Thus, sums over determinants will become multiple integrals.

This properties enabled us in~\cite{IzeKMT99} to obtain the expression of the
matrix elements of the Gaudin matrix~\eqref{matrix-gaudin} in the thermodynamic
limit:
\begin{alignat}{2}
   \Phi_{a b}'(\alpha)&=-2i\pi M \big\{\delta_{a b}\rho_{\text{tot}}(\alpha_a)
                        +\frac 1M  K(\alpha_a-\alpha_b)\big\} +O(M^{-\infty})
                              \quad &\text{for}&\ -1<\Delta<1,\label{gaudinlt-}\\ 
   \Phi_{a b}'(\alpha)&=2\pi M \big\{\delta_{a b}\rho_{\text{tot}}(\alpha_a)
                        +\frac 1M  K(\alpha_a-\alpha_b)\big\}+O(M^{-\infty})
                              \quad &\text{for}&\quad \Delta>1.
\label{gaudinlt}
\end{alignat}
These expressions  will be useful in the following to compute the determinants which appear
in the formulae for the correlation functions in the thermodynamic limit.
Finally, we will obtain correlation functions as multiple integrals of usual
functions of the model.  In the next section we describe in details the computation 
of the simplest $m$-point correlation function using this method.


\section{Emptiness formation probability}
\setcounter{equation}{0}

We consider now the simplest $m$-point correlation function: the emptiness
formation probability, i.e. the probability to detect a ferromagnetic
domain of length $m$ in the antiferromagnetic ground state of the
XXZ model. This probability can be expressed in the following form:

\begin{equation}
\tau(m)=\frac {\bra{\psi_g}\pl_{j=1}^m 
\frac 12(1-\sigma^z_j)\ket{\psi_g}}{\bra{\psi_g}\psi_g\r},
\end{equation}
where $\ket{\psi_g}$ is the ground state in the massless  case and any one of two
ground states constructed by the algebraic Bethe ansatz in the massive regime.

Using the solution of the quantum inverse scattering problem (\ref{szA-D}) one can express
the operators $\frac 12(1-\sigma^z_j)$ in terms of the monodromy matrix elements:
\[\frac 12(1-\sigma^z_j)=\pl_{k=1}^{j-1}\Bigl(A(\xi_k)+D(\xi_k)\Bigr)
\,D(\xi_j)\,\pl_{k=j+1}^{M}\Bigl(A(\xi_k)+D(\xi_k)\Bigr).\]
So the emptiness formation probability can be written uniquely in terms of the monodromy 
matrix elements:
\begin{equation}
\tau(m)=\phi_m(\{\la\})
\frac{\bra{0}\pl_{a=1}^N C(\la_a)\pl_{j=1}^m D(\xi_j)\pl_{a=1}^N B(\la_a)\ket{0}}
{\bra{0}\pl_{a=1}^N C(\la_a)\pl_{a=1}^N B(\la_a)\ket{0}},
\label{efpddd}
\end{equation}
where $\phi_m(\{\la\})$ is the ground state eigenvalue of the corresponding
product of the transfer matrices:
\[\phi_m(\{\la\})=\pl_{j=1}^m\pl_{a=1}^N\frac{\sinh(\la_a-\xi_j)}{\sinh(\la_a-\xi_j+\eta)}.\]

Using the relation (\ref{dbbb}) we obtain the following
 action of a product  of the operators $D(\la)$ on  a state constructed by 
the action of the operators $C(\la)$, 
\begin{equation}
\bra{0}\pl_{k=1}^N C(\la_k)\pl_{j=1}^m
D(\la_{N+j})=\sul_{a_1=1}^{N+1}
\sul_{a_{2}=1\atop{a_{2}\neq a_1}}^{N+2}\dots\!\!
\sul_{a_{m}=1\atop{a_{m}\neq a_1,\dots,a_{m-1}}}^{N+m}\!\!G_{a_1\dots a_m}(\la_1\dots\la_{N+m})
\bra{0}\pl_{k=1\atop{k\neq a_{1},\dots,a_m}}^{N+m} C(\la_k),
\label{efpsum}
\end{equation}
where the function $G$ can be written as:
\begin{equation}
G_{a_1\dots a_m}(\la_1,\dots\,la_{N+m})=\pl_{j=1}^m 
d(\la_{a_j})
\frac{\pl_{b=1\atop{b\neq a_{1},\dots,a_{j-1}}}^{N+j-1}\sinh(\la_{a_j}-\la_b+\eta)}
{\pl_{b=1\atop{b\neq a_1,\dots,a_j}}^{N+j}
          \sinh(\la_{a_j}-\la_b)}.
\label{G}
\end{equation}

To compute the emptiness formation probability one should take the parameters $\la_a$
for $a>N$ equal to $\xi_{a-N}$. It means, in particular, that the sums should be taken 
up to $a_j=N$ as $d(\xi_k)=0$. 

Now we calculate the scalar products in each term of the sum. In fact we have to
calculate the following ``normalized'' product:
\begin{equation}
\mathbb{S}(\{\la_1,\dots,\la_{N-m},\xi_1,\dots,\xi_m\},\{\la\})=
\frac{\bra{0}\pl_{b=1}^{N-m}
C(\la_b)\pl_{k=1}^{m}C(\xi_k)\pl_{k=1}^{N}B(\la_k)\ket{0}}
{\bra{0}\pl_{k=1}^{N}C(\la_k)\pl_{k=1}^{N}B(\la_k)\ket{0}}.
\end{equation} 
This quantity can be easily calculated using the representation for the scalar
products (\ref{theorem1}) (as one of the states in the numerator is a Bethe state) and the
Gaudin formula (\ref{gaudin}) for the norm of Bethe vectors.
Finally we obtain,
\begin{align}
\mathbb{S}(\{\la_1,\dots,\la_{N-m},\xi_1,\dots,\xi_m\},\{\la\})=&\!\pl_{j,k=1\atop{j>k}}^m
\frac{\sinh(\la_{N-m+k}-\la_{N-m+j})}{\sinh(\xi_k-\xi_j)}\pl_{j=1}^m\pl_{k=1}^{N-m}
\frac{\sinh(\la_k-\la_{N-m+j})}{\sinh(\la_k-\xi_j)}
\nonumber\\
\times&\pl_{a=1}^N\pl_{k=1}^m
\frac{\sinh(\la_a-\xi_k+\eta)}{\sinh(\la_a-\la_{N-m+k}+\eta)}
\frac{\det\Psi'(\{\la\},\{\xi\})}{\det\Phi'(\{\la\})},
\label{spnorm}
\end{align}
where the $N\times N$ matrix $\Phi'$ is the Gaudin matrix (\ref{matrix-gaudin}). The first $N-m$ columns
of the $N\times N$ matrix  $\Psi'$ are the same as in the Gaudin matrix but
 the other columns are different:
\begin{alignat*}{2}
\Psi'_{a b}&=\Phi'_{a b},\quad &&b\le N-m,\\
\Psi'_{a b}&=\frac{\sinh\eta}{\sinh(\la_a-\xi_{b+m-N})\sinh(\la_a-\xi_{b+m-N}+\eta)},\quad &&b>N-m.
\end{alignat*}
As the Gaudin matrix is invertible the fraction of the two determinants  in (\ref{spnorm}) can be 
represented as one determinant:
\[\frac{\det\Psi'(\{\la\},\{\xi\})}{\det\Phi'(\{\la\})}=
\det({\Phi'}^{-1}(\{\la\})\Psi'(\{\la\},\{\xi\})).\]
The first $N-m$ columns of the matrix ${\Phi'}^{-1}\Psi'$ are the unity matrix 
columns:
\[({\Phi'}^{-1}\Psi')_{a b}=\delta_{a b},\quad b\le N-m.\]
The action of the inverse Gaudin matrix on the other columns of the matrix
$\Psi'$ can be calculated in the thermodynamic limit.
Using the representation for the Gaudin matrix in the thermodynamic
limit (\ref{gaudinlt-}), (\ref{gaudinlt}) and the Lieb equation (\ref{Lieb}) one  concludes that,
\[
\frac 1M\sul_{b=1}^N\Phi'_{a b}\frac{\tilde{\rho}(\la_b-\xi_k+\frac\eta 2)}
{\tilde{\rho}_{\mathrm{tot}}(\la_b)}=\frac{\sinh\eta}{\sinh(\la_a-\xi_k)
\sinh(\la_a-\xi_k+\eta)} 
+O(M^{-\infty}),
\]
where $\tilde{\rho}(\la)$ is defined differently for two regimes:
\begin{alignat*}{2}
\tilde{\rho}(\la)=&\rho(\la),\quad &-1<&\Delta\le 1,\\
\tilde{\rho}(\la)=&i\rho(i\la),\quad &&\Delta> 1.
\end{alignat*}
Thus for the corresponding matrix elements we obtain
\[({\Phi'}^{-1}\Psi')_{a b}=\frac{\tilde{\rho}(\la_a-\xi_{b+m-N}+\frac\eta 2)}
{M\tilde{\rho}_{\mathrm{tot}}(\la_a)}+O(M^{-\infty}),\quad b>N-m.\]
Finally the fraction of the two determinants   in (\ref{spnorm}) can be written
in a very simple form in the thermodynamic limit:
\begin{equation}
\frac{\det\Psi'(\{\la\},\{\xi\})}{\det\Phi'(\{\la\})}=\frac 1{M^m}\pl_{a=1}^m
\tilde{\rho}^{-1}_{\mathrm{tot}}(\la_{N-m+a})\det S(\{\la_{N-m+1},\dots,\la_N\},\{\xi\})
+O(M^{-\infty}),
\label{spS}
\end{equation}
where the $m\times m$ matrix $S$ is:
\begin{equation}
S_{a b}=\tilde{\rho}(\la_{N-m+a}-\xi_{b}+\frac\eta 2).
\label{S}
\end{equation}

Now, using (\ref{efpddd}), (\ref{efpsum}), (\ref{G}), (\ref{spnorm}) and (\ref{spS}),
we obtain the following representation for the emptiness formation probability: 
\begin{equation}
\tau(m)=\frac{1}{M^m\pl_{k<l}\sinh(\xi_k-\xi_l)}\sul_{a_m=1}^{N}
\sul_{a_{m-1}=1}^{N}\dots
\sul_{a_{1}=1}^{N} H(\{\la_{a_1},\dots,\la_{a_m}\},\{\xi\})\pl_{j=1}^m
\tilde{\rho}^{-1}_{\mathrm{tot}}(\la_{a_j}),
\end{equation}
where the function $H$ is:
\begin{align}
H(\{\la_{a_1},\dots,\la_{a_m}\},\{\xi\})=&
\frac{1}{\pl_{k>l}
\sinh(\la_{a_k}-\la_{a_l}+\eta)}\det S(\{\la_{a_1},\dots\la_{a_m}\},\{\xi\})\times\nonumber\\
\times&\pl_{j=1}^m\left(\pl_{k=1}^{j-1}
\sinh(\la_{a_j}-\xi_k+\eta)\pl_{k=j+1}^{m}\sinh(\la_{a_j}-\xi_k)\right)+O(M^{-\infty}).
\end{align}
Here we take the sums over all the values of $a_j$, as if two indices coincide
($a_j=a_k,\,\, j\neq k$) the determinant of the matrix $S$ vanishes.

In the thermodynamic limit for the ground state the sums can be replaced by the integrals (\ref{sumint})
and the emptiness formation probability can be expressed as a multiple integral,
\begin{equation}
\tau(m)=\frac{1}{\pl_{k<l}\sinh(\xi_k-\xi_l)}\int\limits_{-\tilde{\Lambda}}^{\tilde{\Lambda}}d\la_1\dots
\int\limits_{-\tilde{\Lambda}}^{\tilde{\Lambda}}d\la_m H(\{\la_1,\dots,\la_m\},\{\xi\}),
\end{equation}
where $\tilde{\Lambda}=\Lambda$ for $-1<\Delta\le 1$ and  $\tilde{\Lambda}=-i\Lambda$ for $\Delta> 1$.

Thus we obtain explicit results for both regimes of the XXZ model. In both cases
the determinant of the matrix $S$ can be calculated explicitly.
In the massless 
case it is  the Cauchy determinant:
\begin{equation}
\det S=\left(\frac i{2\zeta}\right)^m\frac{\pl_{k<l}\sinh\frac\pi\zeta(\xi_k-\xi_l)\pl_{a>b}
\sinh\frac\pi\zeta(\la_a-\la_b)}{\pl_{a=1}^m\pl_{k=1}^m\sinh\frac\pi\zeta(\la_a-\xi_k)}.
\end{equation}
The emptiness formation probability has in this case the following form:
\begin{align}
\tau(m)=&\pl_{k<l}\frac{\sinh\frac\pi\zeta(\xi_k-\xi_l)}{\sinh(\xi_k-\xi_l)}
\int\limits_{-\infty}^{\infty}  i\frac{d\la_1}{2\zeta}\dots
\int\limits_{-\infty}^{\infty}  i\frac{d\la_m}{2\zeta}
\pl_{a>b}\frac{\sinh\frac\pi\zeta(\la_a-\la_b)}{\sinh(\la_a-\la_b-i\zeta)}
\times\nonumber\\
\times&
\pl_{a=1}^m\pl_{k=1}^m\frac 1{\sinh\frac\pi\zeta(\la_a-\xi_k)}
\pl_{j=1}^m\left(\pl_{k=1}^{j-1}
\sinh(\la_j-\xi_k-i\zeta)\pl_{k=j+1}^m\sinh(\la_j-\xi_k)\right).
\end{align}
In the homogeneous limit ($\xi_j=-i\zeta/2,\, \forall j $) we obtain the following 
result for the emptiness
formation probability:
\begin{align}
\tau(m)=&
(-1)^m\left(-
\frac\pi\zeta
\right)^{\frac{m(m+1)}2}
\int\limits_{-\infty}^{\infty}  \frac{d\la_1}{2\pi}\dots
\int\limits_{-\infty}^{\infty}  \frac{d\la_m}{2\pi}
\pl_{a>b}\frac{\sinh\frac\pi\zeta(\la_a-\la_b)}{\sinh(\la_a-\la_b-i\zeta)}\times\nonumber\\
\times&
\pl_{j=1}^m\frac {\sinh^{j-1}(\la_j-i\frac\zeta 2)\sinh^{m-j}(\la_j+i\frac\zeta 2)}
{\cosh^m\frac\pi\zeta\la_j}.
\end{align}

In the massive case the determinant of the matrix $S$ is more complicated but can be
expressed in terms of the Theta functions: 
\begin{equation}
\det S= g_m\Bigl(-\frac{1}{2\pi}\Bigr)^m 
     \, \frac{\pl_{j<k} \th_1 (i\la_j-i\la_k)\ \th_1 (i\xi_k-i\xi_j)}
             {\pl_{j,k=1}^m \th_1 (i\la_j-i\xi_k)}\ 
              \th_2\biggl(\sul_{j=1}^m(i\la_j-i\xi_j)\biggr), 
\label{Smass}
\end{equation} 
where
\[g_m=\pl_{n=1}^{\infty} \Bigl(\frac{1-q^{2n}}{1+q^{2n}}\Bigr)^2
      \Big[2q^{1/4} \pl_{n=1}^{\infty} (1-q^{2n})^3\Big]^{m-1}.\]
We give the proof of this formula in Appendix A.
As usually in this regime we change the variables for more convenient ones:
$\b=i\xi$, $\zeta=-\eta$. For the emptiness formation probability we obtain:
\begin{align}
\tau(m)=&
g_m\pl_{k<l}\frac{\th_1(\b_k-\b_l)}{\sin(\b_k-\b_l)}
\int\limits_{-\pi/2}^{\pi/2}i\frac{d\la_1}{2\pi }\dots
\int\limits_{-\pi/2}^{\pi/2}i\frac{d\la_m}{2\pi }\,\th_2\left(\sul_{j=1}^m(\la_j-\b_j)\right)
\pl_{a>b}\frac{\th_1(\la_a-\la_b)}{\sin(\la_a-\la_b-i\zeta)}
\times\nonumber\\
\times&
\pl_{a=1}^m \left(\pl_{k=1}^m\frac 1{\th_1(\la_a-\b_k)}
\pl_{k=1}^{a-1}
\sin(\la_a-\b_k-i\zeta)\pl_{k=a+1}^m\sin(\la_a-\b_k)\right).
\end{align}
In the homogenous limit $\b_j=-i\zeta/2$ we have the following result:
\begin{align}
\tau(m)=&
\pl_{n=1}^{\infty} \Bigl(\frac{1-q^{2n}}{1+q^{2n}}\Bigr)^2
      \Big[2q^{1/4} \pl_{n=1}^{\infty} (1-q^{2n})^3\Big]^{\frac{m(m+1)}2-1}
\int\limits_{-\pi/2}^{\pi/2}i\frac{d\la_1}{2\pi }\dots
\int\limits_{-\pi/2}^{\pi/2}i\frac{d\la_m}{2\pi }\,
\times\nonumber\\
\times&
\th_2\left(\sul_{j=1}^m(\la_j+i\frac\zeta 2)\right)
\pl_{a>b}\frac{\th_1(\la_a-\la_b)}{\sin(\la_a-\la_b-i\zeta)}
\pl_{j=1}^m\frac {\sin^{j-1}(\la_j-i\frac\zeta 2)\sin^{m-j}(\la_j+i\frac\zeta 2)}
{\th_1^m(\la_j+i\frac\zeta 2)}.
\end{align}

We have shown in this section that the simplest $m$-point correlation function for the XXZ 
model can be expressed using the algebraic Bethe ansatz 
as multiple integrals of the elementary or elliptic functions. These results
reproduce for this particular case the formulae obtained by Jimbo and Miwa \cite{JimM95L,JimMMN92,JimM96}.
In the isotropic XXX limit ($\Delta=1$) we also obtain a formula derived in \cite{Nak94,KorIEU94} (using the 
 Jimbo and Miwa method).
In the
next section we apply the same approach to calculate more general correlation functions of the XXZ
chain.


\section{Correlation functions}
\setcounter{equation}{0}

In this section we consider a more general case of correlation functions : the ground
state mean value 
of any   product of the local elementary $2\times 2$ matrices 
$E^{\e',\e}_{l k}=\delta_{l,\e'}\delta_{k,\e}$: 
\begin{equation}
F_m(\{\e_j,\e'_j\})=\frac {\bra{\psi_g}\pl_{j=1}^m 
E^{\e'_j,\e_j}_j\ket{\psi_g}}{\bra{\psi_g}\psi_g\r} .
\label{genabcd0}
\end{equation}
It should be mentioned that an arbitrary $n$-point correlation function can be obtained 
as a sum of such mean values.

To calculate this product we use at first the solution of the quantum inverse 
scattering problem (\ref{s-B})-(\ref{szA-D}),
 representing the local elementary matrices in terms of the corresponding
monodromy matrix elements:
\[E^{\e'_j,\e_j}_j=\pl_{k=1}^{j-1}\Bigl(A(\xi_k)+D(\xi_k)\Bigr)
T_{\e_j,\e'_j}(\xi_j)\pl_{k=j+1}^{M}\Bigl(A(\xi_k)+D(\xi_k)\Bigr).\]

Thus we reduce the problem to the computation of the ground state
 mean value of an arbitrary ordered product of the 
monodromy matrix elements,
\begin{equation}
F_m(\{\e_j,\e'_j\})=\phi_m(\{\la\})\frac {\bra{\psi_g}
T_{\e_1,\e'_1}(\xi_1)\dots T_{\e_m,\e'_m}(\xi_m)\ket{\psi_g}}{\bra{\psi_g}\psi_g\r}, 
\label{genabcd}
\end{equation}
where $\phi_m(\{\la\})$ is the ground state eigenvalue of the corresponding
product of the transfer matrices:
\[\phi_m(\{\la\})=\pl_{j=1}^m\pl_{a=1}^N\frac{\sinh(\la_a-\xi_j)}{\sinh(\la_a-\xi_j+\eta)}.\]

Now to calculate these mean values we use the commutation relations
of the monodromy matrix elements.

An arbitrary product of the monodromy matrix elements can be treated in a rather
general way. At first one should consider the two following sets of indices:
\begin{alignat*}{4}
&\mathbf{\a^+}=\{j:\, 1\le j\le m, \,\e_j=1\},&\quad&\mathrm{card}(\mathbf{\a^+})=s',
\quad&\mathrm{max}_{j\in\mathbf{\a^+}}(j)\equiv j'_{\mathrm{max}},
\quad&\mathrm{min}_{j\in\mathbf{\a^+}}(j)\equiv j'_{\mathrm{min}},\\ 
&\mathbf{\a^-}=\{j:\, 1\le j\le m,\, \e'_j=2\},&\quad&\mathrm{card}(\mathbf{\a^-})=s,
\quad&\mathrm{max}_{j\in\mathbf{\a^-}}(j)\equiv j_{\mathrm{max}},
\quad&\mathrm{min}_{j\in\mathbf{\a^-}}(j)\equiv j_{\mathrm{min}}.
\end{alignat*}
It should be mentioned that in  a 
general case the intersection of these two sets 
is not empty and corresponds to the operators $B(\xi_j)$.

Consider now the action of an arbitrary product on a state
constructed by the action of the operators $C(\la)$,
\[
\bra{0}\pl_{k=1}^N C(\la_k)T_{\e_1,\e'_1}(\la_{N+1})\dots
T_{\e_m,\e'_m}(\la_{N+m}),
\]
applying one by one
the formulae (\ref{abbb})-(\ref{cbbb}). For all the indices $j$ from the
sets $\mathbf{\a^+}$ and $\mathbf{\a^-}$ one obtains a summation on the 
corresponding indices $a'_j$ (for $j\in\mathbf{\a^+}$, corresponding to the
action of the operators $A(\la)$ or $B(\la)$) or $a_j$ 
(for $j\in\mathbf{\a^-}$,  corresponding to the
action of the operators $D(\la)$ or $B(\la)$).  As the product of the monodromy matrix elements
is ordered these summations are also ordered and the corresponding indices 
should be taken from the following sets:
\begin{align*}
\mathbf{A}_j=&\{b: 1\le b\le N+m, \,\,b\neq a_k,a'_k,\,\, k<j\},\\
\mathbf{A'}_j=&\{b: 1\le b\le N+m, \,\,b\neq a'_k, \,k<j,\, 
b\neq a_k, \,k\le j\}.
\end{align*}
Thus the action of a product of the monodromy matrix elements 
can be written as the following sum:
\begin{equation}
\bra{0}\pl_{k=1}^N C(\la_k)\,T_{\e_1,\e'_1}(\la_{N+1})\dots
T_{\e_m,\e'_m}(\la_{N+m})=
\sul_{\{a_j,a'_j\}}G_{\{a_j,a'_j\}}(\la_1,\dots,\la_{N+m})\bra{0}\pl_{b\in \mathbf{A}_{m+1}}
C(\la_b)
\label{action}
\end{equation}
The summation is taken over the indices $a_j$ for  $j\in\mathbf{\a^-}$ and $a'_j$ for 
 $j\in\mathbf{\a^+}$ such that:
\[1\le a_j\le N+j, \,\, 
a_j\in \mathbf{A}_j,\quad 1\le a'_j\le N+j, \,\, a'_j\in \mathbf{A'}_j.\]
The functions $G_{\{a_j,a'_j\}}(\la_1,\dots\la_{N+m})$ can be easily obtained
from the formulae (\ref{abbb})-(\ref{cbbb}) taking into acount that $\la_{a}=\xi_{N-a}$
for $a>N$:
\begin{align}
G_{\{a_j,a'_j\}}(\la_1,\dots,\la_{N+m})=&\pl_{j\in\mathbf{\a^-}}d(\la_{a_j})
\frac{\pl_{b=1\atop{b\in\mathbf{A}_j}}^{N+j-1}\sinh(\la_{a_j}-\la_b+\eta)}
{\pl_{b=1\atop{b\in\mathbf{A'}_j}}^{N+j}
          \sinh(\la_{a_j}-\la_b)}\times\nonumber\\
                                 \times&\pl_{j\in\mathbf{\a^+}}
                 a(\la_{a'_j})
\frac{\pl_{b=1\atop{b\in\mathbf{A'}_j}}^{N+j-1                                            
                                                }\sinh(\la_b-\la_{a'_j}+\eta)}
{\pl_{b=1\atop{b\in\mathbf{A}_{j+1}}}^{N+j}\sinh(\la_b-\la_{a'_j})}.
\label{funabcd}
\end{align}

Now to calculate the normalized mean value (\ref{genabcd}) we apply the 
representation for the scalar product (\ref{theorem1})
and the Gaudin formula (\ref{gaudin}).
It should be mentioned that the number of operators $C(\la)$ has to be
equal to the number of the operators $B(\la)$, as otherwise the mean value is zero, and
hence the total number of elements in the sets $\mathbf{\a^+}$ and $\mathbf{\a^-}$ is
 $s+s'=m$.
Taking into account that in (\ref{genabcd}), for $b>N,\,\, \la_b=\xi_{b-N}$ one can 
consider the scalar products appearing in the representation for the ground state mean values, 
\[\frac{\bra{0}\pl_{b\in \mathbf{A}_{m+1}}
C(\la_b)\pl_{k=1}^{N}B(\la_k)\ket{0}}{\bra{0}\pl_{k=1}^{N}C(\la_k)\pl_{k=1}^{N}B(\la_k)\ket{0}},\]
for all the permitted values of $a_j, a'_j$ 
using the same method as for the emptiness formation probability. Finally 
we obtain:
\begin{equation}
F_m(\{\e_j,\e'_j\})=\frac{1}{\pl_{k<l}\sinh(\xi_k-\xi_l)}
\sul_{\{a_j,a'_j\}}H_{\{a_j,a'_j\}}(\la_1,\dots,\la_{N+m}),
\label{sum1}
\end{equation}
the sum being taken on the same set of indices $a_j,a'_j$ as in (\ref{action}). The functions
$H_{\{a_j,a'_j\}}(\{\la\})$ can be obtained using (\ref{funabcd}) and the
representations for the scalar products. It is convenient  to introduce the following 
set of indices:
\[\{b_1,\dots,b_m\}=\{a'_{j'_{\mathrm{max}}},\dots,a'_{j'_{\mathrm{min}}},
a_{j_{\mathrm{min}}},\dots,a_{j_{\mathrm{max}}}\}.\]
One should also take into account that for the XXZ model $a(\la)=1$ and $d(\xi_k)=0$.
Then one obtains that $a_j\le N$, $\forall j\in\mathbf{\a^-}$, (otherwise the corresponding
term is zero):
\begin{align}   
H_{\{a_j,a'_j\}}(\{\la\})=&
\frac{(-1)^{s'} }{\pl_{k>l}
\sinh(\la_{b_k}-\la_{b_l}+\eta)}\pl_{j\in\mathbf{\a^+}}\left(\pl_{k=1}^{j-1}\!\!
\sinh(\la_{a'_j}-\xi_k-\eta)\pl_{k=j+1}^{m}\sinh(\la_{a'_j}-\xi_k)\right)\times\nonumber\\
\times&
\pl_{j\in\mathbf{\a^-}}\!\!\left(\pl_{k=1}^{j-1}
\sinh(\la_{a_j}-\xi_k+\eta)\pl_{k=j+1}^{m}\sinh(\la_{a_j}-\xi_k)\right)\!\!
\biggl(\det M(\{b_k\})+O(M^{-\infty})\biggr),
\label{underint}
\end{align}
where the $m\times m$ matrix $M(\{b_k\})$ is slightly different in comparison to  the 
case of the emptiness 
formation probability,
\begin{align*}
b_l>N,\quad M_{l k}&=-\delta_{b_l-N,k},\\
b_l\le N,\quad M_{l k}&=\frac{\tilde{\rho}(\la_{b_l}-\xi_k+\frac\eta 2)}
{\tilde{\rho}_{\mathrm{tot}}(\la_{b_l})}.
\end{align*}
The sum in (\ref{sum1}) can be rewritten in a more simple way if one takes into account
that the function $H_{\{a_j,a'_j\}}(\{\la\})$ defined by (\ref{underint}) is equal to zero
if $b_j=b_k,\,\, j\neq k$ (as the determinant vanishes in this case), or if $a'_j>N+j$:
\begin{equation}
F_m(\{\e_j,\e'_j\})=\frac{1}{\pl_{k<l}\sinh(\xi_k-\xi_l)}
\sul_{b_1=1}^{N+m}\dots\sul_{b_{s'}=1}^{N+m}
\sul_{b_{s'+1}=1}^{N}
\dots\sul_{b_m=1}^N H_{\{a_j,a'_j\}}(\{\la\}).
\end{equation} 
The sum over $1\le b_j\le N$ is just a sum over the rapidities in the ground state and
can be replaced by  integrals as in the case of the emptiness formation probability:
\[\sul_{a=1}^N f(\la_a)=\int\limits_{-\tilde{\Lambda}}^{\tilde{\Lambda}}d\la\, \tilde{\rho}_{\mathrm{tot}}(\la)f(\la)
+O(M^{-\infty})
 .\] 
The
contributions of the terms with $a'_j>N$ can be rewritten as   integrals over the 
contours $\Gamma_{a'_j-N}$ surrounding the
 pole of the corresponding density function $\tilde{\rho}(\la-\xi_{a'_j-N}+\frac\eta 2)$ in the
point $\la=\xi_{a'_j-N}$. The residues of the density function $\rho(\la-\xi+\frac\eta 2)$ (for both regimes)
 in this 
points are:
\[2\pi i \left.\Res\biggl(\rho(\la-\xi+\frac\eta 2)\biggr)\right|_{\la=\xi}=-1.\]
The other points $\xi_l$ should be outside
the  contour $\Gamma_{a'_j-N}$. The matrix $M(\{b_k\})$ then should
be replaced by the same matrix $S(\{\la\},\{\xi\})$ as
in the representation for the emptiness formation probability.
 Finally for the correlation function one obtains:
\begin{align}
F_m(\{\e_j,\e'_j\})=\frac{1}{\pl_{k<l}\sinh(\xi_k-\xi_l)}
&\left( \int\limits_{-\tilde{\Lambda}}^{\tilde{\Lambda}}+\sul_{j=1}^m\oint_{\Gamma_j}\right)d\la_1\dots
\left( \int\limits_{-\tilde{\Lambda}}^{\tilde{\Lambda}}+\sul_{j=1}^m\oint_{\Gamma_j}\right)d\la_{s'} 
\times\nonumber\\
&\int\limits_{-\tilde{\Lambda}}^{\tilde{\Lambda}}d\la_{s'+1}\dots 
\int\limits_{-\tilde{\Lambda}}^{\tilde{\Lambda}}d\la_{m} 
\tilde{H}_{\{\e_j,\e'_j\}}(\la_1,\dots,\la_m)+O(M^{-\infty}), 
\label{intabcd1}
\end{align}
where the function $\tilde{H}_{\{\e_j,\e'_j\}}(\la_1,\dots,\la_m)$ is defined as,
\begin{align}   
\tilde{H}_{\{\e_j,\e'_j\}}(\{\la\})=&
\frac{(-1)^{s'} }{\pl_{k>l}
\sinh(\la_k-\la_l+\eta)}\pl_{j\in\mathbf{\a^-}}\left(\pl_{k=1}^{j-1}
\sinh(\mu_j-\xi_k+\eta)\pl_{k=j+1}^{m}\sinh(\mu_j-\xi_k)\right)\times\nonumber\\
\times&\pl_{j\in\mathbf{\a^+}}\left(\pl_{k=1}^{j-1}
\sinh(\mu'_j-\xi_k-\eta)\pl_{k=j+1}^{m}\sinh(\mu'_j-\xi_k)\right)
\det S(\{\la\},\{\xi\}),
\label{underint1}
\end{align}
where
\[S_{lk}=\tilde{\rho}(\la_l-\xi_k+\frac\eta 2),\]
and the parameters of integration are ordered in the following way:
\[\{\la_1,\dots\la_m\}=\{\mu'_{j'_{\mathrm{max}}},\dots,\mu'_{j'_{\mathrm{min}}},
\mu_{j_{\mathrm{min}}},\dots,\mu_{j_{\mathrm{max}}}\}.\]

Consider now separately the two regimes of the XXZ model. In the massless regime
$\eta=-i\zeta$ is imaginary, the ground state rapidities $\la$ are real and the limit
of integration is infinity $\Lambda=\infty$. In this case we consider the inhomogeneity
parameters $\xi_j$ such that $0>\mathrm{Im}(\xi_j)>-\zeta$. The function 
 $\tilde{H}_{\{\e_j,\e'_j\}}(\la_1,\dots,\la_m)$  for all the arguments $\la_j$ in
the region $0>\mathrm{Im}(\la_j)>-\zeta$
has only simple poles in the points $\la_j=\xi_k$. Hence the sums of  integrals
in    (\ref{intabcd1}) can be rewritten 
as one integral on a displaced contour:
\[\left( \int\limits_{-\infty}^{\infty}+\sul_{j=1}^m\oint_{\Gamma_j}\right)d\la_j
\longrightarrow\int\limits_{-\infty-i\zeta}^{\infty-i\zeta}d\la_j.\]

Finally for the correlation functions in the thermodynamic limit
 one obtains the following result in this regime:  
\begin{align}
F_m(\{\e_j,\e'_j\})=\frac{1}{\pl_{k<l}\sinh(\xi_k-\xi_l)}
\int\limits_{-\infty-i\zeta}^{\infty-i\zeta}\!\!\!\!\!d\la_1\dots\!\!\!\!
\int\limits_{-\infty-i\zeta}^{\infty-i\zeta}\!\!\!d\la_{s'} 
\int\limits_{-\infty}^{\infty}d\la_{s'+1}\dots \int\limits_{-\infty}^{\infty}d\la_{m} 
\tilde{H}_{\{\e_j,\e'_j\}}(\la_1,\dots,\la_m). 
\label{intabcd2}
\end{align}

Now one can rewrite this result   using the corresponding
representations for the determinants of the matrix $S(\{\la\},\{\xi\})$:
\begin{align}
F_m(\{\e_j,\e'_j\})=&
\pl_{k<l}\frac{\sinh\frac\pi\zeta(\xi_k-\xi_l)}{\sinh(\xi_k-\xi_l)}
\pl_{j=1}^{s'}\int\limits_{-\infty-i\zeta}^{\infty-i\zeta} \frac{d\la_j}{2i\zeta}
\pl_{j=s'+1}^{m}\int\limits_{-\infty}^{\infty}i\frac{d\la_j}{2\zeta}
\pl_{a>b}\frac{\sinh\frac\pi\zeta(\la_a-\la_b)}{\sinh(\la_a-\la_b-i\zeta)}
\times\nonumber\\
\times&
\pl_{a=1}^m\pl_{k=1}^m\frac 1{\sinh\frac\pi\zeta(\la_a-\xi_k)}
\pl_{j\in\mathbf{\a^-}}\left(\pl_{k=1}^{j-1}
\sinh(\mu_j-\xi_k-i\zeta)\pl_{k=j+1}^m\sinh(\mu_j-\xi_k)\right)\times\nonumber\\
\times&\pl_{j\in\mathbf{\a^+}}\left(\pl_{k=1}^{j-1}
\sinh(\mu'_j-\xi_k+i\zeta)\pl_{k=j+1}^m\sinh(\mu'_j-\xi_k)\right).
\end{align}

In the homogeneous limit ($\xi_j=-i\zeta/2,\, \forall j $) the correlation function $F_m(\{\e_j,\e'_j\})$ 
has the following form:
\begin{align}
F_m(\{\e_j,\e'_j\})=&
(-1)^{s}\left(-
\frac\pi\zeta
\right)^{\frac{m(m+1)}2}
\pl_{j=1}^{s'}\int\limits_{-\infty-i\zeta}^{\infty-i\zeta}  \frac{d\la_j}{2\pi}
\pl_{j=s'+1}^{m}\int\limits_{-\infty}^{\infty}\frac{d\la_j}{2\pi}
\pl_{a>b}\frac{\sinh\frac\pi\zeta(\la_a-\la_b)}{\sinh(\la_a-\la_b-i\zeta)}\times\nonumber\\
\times&
\pl_{j\in\mathbf{\a^-}}\frac {\sinh^{j-1}(\mu_j-i\frac\zeta 2)\sinh^{m-j}(\mu_j+i\frac\zeta 2)}
{\cosh^m\frac\pi\zeta\mu_j}\times\nonumber\\
\times&
\pl_{j\in\mathbf{\a^+}}\frac{\sinh^{j-1}(\mu'_j+3i\frac\zeta 2)
\sinh^{m-j}(\mu'_j+i\frac\zeta 2)}{\cosh^m\frac\pi\zeta\mu'_j}.
\end{align}   
These results agree exactly with the ones obtained
by Jimbo and Miwa in \cite{JimM96}, taking into account that their Hamiltonian is obtained from our one by the transformation $UH_\Delta U^{-1}=-H_{-\Delta}$, $U=\prod_{j=1}^{M \over 2}\s_{2j}^z$. 

Similarly in the massive regime the parameter $\eta=-\zeta$ is real, the solutions
of the Bethe equations corresponding to the two ground states are imaginary $\la_a=-i\a_a$,  and
$\tilde{\Lambda}=-i\pi/2$. The
inhomogeneity   parameters are chosen in such a way that $0>\mathrm{Im}(\b_j)>-\zeta $,
$\xi=-i\beta$. 
Taking into account that the function $\tilde{H}_{\{\e_j,\e'_j\}}(\la_1,\dots,\la_m)$
is $\pi$-periodic for all the arguments $\a_a$ and that  in the region 
$0>\mathrm{Im}(\a_a)>-\zeta$, $-\pi/2\le\mathrm{Re}(\a_a)\le\pi/2$, it has 
only simple poles  in the points $\a_a=\b_k$ one
can rewrite the sums of integrals in the same way as in the massless situation:
\[\left( \int\limits_{-\pi/2}^{\pi/2}+\sul_{j=1}^m\oint_{\Gamma_j}\right)d\a_j
\longrightarrow\int\limits_{-\pi/2-i\zeta}^{\pi/2-i\zeta}d\a_j.\]
Finally using the formula for the determinant of the matrix $S$ (\ref{Smass}) in this case
one obtains:
\begin{align}
F_m(\{\e_j,\e'_j\})=&
g_m\pl_{k<l}\frac{\th_1(\b_k-\b_l)}{\sin(\b_k-\b_l)}
\pl_{j=1}^{s'}\int\limits_{-\pi/2-i\zeta}^{\pi/2-i\zeta} \frac{d\la_j}{2\pi i}
\pl_{j=s'+1}^{m}\int\limits_{-\pi/2}^{\pi/2}i\frac{d\la_j}{2\pi }
\pl_{a>b}\frac{\th_1(\la_a-\la_b)}{\sin(\la_a-\la_b-i\zeta)}
\times\nonumber\\
\times&
\pl_{a=1}^m\pl_{k=1}^m\frac 1{\th_1(\la_a-\b_k)}
\pl_{j\in\mathbf{\a^-}}\left(\pl_{k=1}^{j-1}
\sin(\mu_j-\b_k-i\zeta)\pl_{k=j+1}^m\sin(\mu_j-\b_k)\right)\times\nonumber\\
\times&\pl_{j\in\mathbf{\a^+}}\left(\pl_{k=1}^{j-1}
\sin(\mu'_j-\b_k+i\zeta)\pl_{k=j+1}^m\sin(\mu'_j-\b_k)\right)\th_2\left(\sul_{j=1}^m(\la_j-\b_j)\right),
\end{align}
where $\{\la_1,\dots\la_m\}=\{\mu'_{j'_{\mathrm{max}}},\dots,\mu'_{j'_{\mathrm{min}}},
\mu_{j_{\mathrm{min}}},\dots,\mu_{j_{\mathrm{max}}}\}$ and
\[g_m=\pl_{n=1}^{\infty} \Bigl(\frac{1-q^{2n}}{1+q^{2n}}\Bigr)^2
      \Big[2q^{1/4} \pl_{n=1}^{\infty} (1-q^{2n})^3\Big]^{m-1}.\]
This result is equivalent to the formula \cite{JimMMN92,JimM95L} obtained as a solution
of the $q$-KZ equations.

In the homogenous limit $\b_j=-i\zeta/2$ we obtain the following result for the correlation function:
\begin{align}
F_m(\{\e_j,\e'_j\})=&
\pl_{n=1}^{\infty} \Bigl(\frac{1-q^{2n}}{1+q^{2n}}\Bigr)^2
      \Big[2q^{1/4} \pl_{n=1}^{\infty} (1-q^{2n})^3\Big]^{\frac{m(m+1)}2-1}
\pl_{j=1}^{s'}\!\!\int\limits_{-\pi/2-i\zeta}^{\pi/2-i\zeta}  \frac{d\la_j}{2\pi i}
\pl_{j=s'+1}^{m}\int\limits_{-\pi/2}^{\pi/2}i\frac{d\la_j}{2\pi }\times\nonumber\\
\times&
\pl_{a>b}\frac{\th_1(\la_a-\la_b)}{\sin(\la_a-\la_b-i\zeta)}
\pl_{j\in\mathbf{\a^-}}\frac {\sin^{j-1}(\mu_j-i\frac\zeta 2)\sin^{m-j}(\mu_j+i\frac\zeta 2)}
{\th_1^m(\mu_j+i\frac\zeta 2)}\times\nonumber\\
\times&
\pl_{j\in\mathbf{\a^+}}\frac{\sin^{j-1}(\mu'_j+3i\frac\zeta 2)
\sin^{m-j}(\mu'_j+i\frac\zeta 2)}{\th_1^m(\mu'_j+i\frac\zeta 2)}
\th_2\left(\sul_{j=1}^m(\la_j+i\frac\zeta 2)\right).
\end{align}

Thus for both regimes the results for the correlation functions obtained by 
Jimbo, Miwa and collaborators \cite{JimMMN92,JimM95L,JimM96}
as solutions of the $q$-KZ equation can be reproduced using the algebraic Bethe ansatz approach.
We have shown also that the main difference between the two regimes is the
determinant of the matrix the elements of which are the density function taken at the corresponding
 values of the spectral parameter. In both cases this determinant can be calculated
in terms of elementary or elliptic functions, which leads to the different 
representations for the two regimes.

It should be mentioned that to calculate the correlation functions one can also use
the action of the monodromy matrix elements on a state constructed by the action of the operators
$B(\la)$. It leads to a similar but different representation for the correlation functions.


\section{External magnetic field}

\setcounter{equation}{0}
In this section we show how the  previous results can be generalized 
for the XXZ spin chain in a constant external magnetic field:
\begin{equation}
\mathbf{H}_h= \mathbf{H}_{\mathrm{XXZ}}- h S_z,
\end{equation}
where $S_z$ is the third component of the total spin,
\[S_z=\frac 12 \sul_{j=1}^M\s_j^z.\]
The third component of the total spin commutes with the Hamiltonian of the XXZ model
and also with the transfer matrix and so the eigenstates remain the same as in
the case of zero magnetic field. However the ground state changes and the 
corresponding Lieb
equation changes also. In the massless regime the ground state changes for any 
value of the magnetic field. The Lieb equation now
has the following form,
\begin{equation}
\rho_h(\la)+\int\limits_{-\Lambda_h}^{\Lambda_h}d\mu K(\la-\mu)\rho_h(\mu)=\frac 1{2\pi}
p'_0(\la),
\label{hlieb}
\end{equation} 
where the Fermi momentum $\Lambda_h$ is defined by the following integral 
equation for the excitation energy:
\begin{equation}
\eps_h(\la)+\int\limits_{-\Lambda_h}^{\Lambda_h}d\mu K(\la-\mu)\eps_h(\mu)=\eps_0(\la),\quad 
\eps_h(\Lambda_h)=0,
\end{equation} 
where the bare energy $\eps_0(\la )$ is,
\[\eps_0=h-2\sin\zeta\, p'_0(\la).\]

In the massive regime the ground state changes
only for the magnetic field greater than the critical value equal to the gap width. Indeed
the equations defining the Fermi momentum,
\begin{equation}
\eps_h(\la)+\int\limits_{-\Lambda_h}^{\Lambda_h}d\mu K(\la-\mu)\eps_h(\mu)=\eps_0(\la),\quad 
\eps_h(\Lambda_h)=0,
\end{equation}
with the bare energy 
\[\eps_0=h-2\sinh\zeta\, p'_0(\la),\]
have no solution for $\Lambda_h\le\pi/2$ if the magnetic field $h<h_c$, where the critical field
is:
\[ h_c=\frac 2\pi\sinh\zeta\sul_{n=-\infty}^{\infty}\frac{(-1)^n}{\cosh\, n\zeta}.\]
So if the magnetic field is under its critical value, the ground states, the density
function $\rho(\la)$ and the zero temperature
correlation functions do not change. If the magnetic field is greater than $h_c$
there is a solution for the Fermi momentum $\Lambda_h$ and the density function is given
by the equation (\ref{hlieb}) with  the kernel $K(\la)$ and the function $p_0'(\la)$ corresponding
to the massive regime. It should be mentioned that in this case there is no more gap in
the spectrum, and the XXZ model with $\Delta>1$ and $h>h_c$ is  massless.

In  general these equations cannot be solved explicitly as in the case of zero magnetic field, but
as the integral operators are rather simple the solution can be obtained numerically with any
given precision.

The function $\rho_h(\la)$ has only one simple pole in the region $0>\mathrm{Im}(\la)>-\zeta$ in the point 
$\la=-i\zeta/2$ and 
its residue
is, 
\[2\pi i \left.\Res\biggl(\rho_h(\la)\biggr)\right|_{\la=-i\frac\zeta 2}=-1.\] 

The correlation functions can be calculated using exactly the same method as in the previous
section. It should be mentioned that in this case only the homogeneous limit is interesting.
 Using the same arguments one can show that in the  regime $|\Delta|<1$ 
the general correlation 
function can be written   as follows,
\begin{align}
F_m(h,\{\e_j,\e'_j\})=&
\pl_{j=1}^{s'}\left(\int\limits_{-\Lambda_h}^{-\infty}+
\int\limits_{-\infty-i\zeta}^{\infty-i\zeta}+\int\limits_{\infty}^{\Lambda_h}\right)d\la_{j} 
\,\pl_{j=s'+1}^m \int\limits_{-\Lambda_h}^{\Lambda_h}d\la_{j} \frac{(-1)^{s'} }{\pl_{k>l}
\sinh(\la_k-\la_l-i\zeta)}\times \nonumber\\
\times&\pl_{j\in\mathbf{\a^-}}\left(
\sinh^{j-1}(\mu_j-i\frac\zeta 2)\sinh^{m-j}(\mu_j+i\frac\zeta 2)\right)\times\nonumber\\
\times&
\pl_{j\in\mathbf{\a^+}}\left(
\sinh^{j-1}(\mu'_j+3i\frac{\zeta} 2)\sinh^{m-j}(\mu'_j+i\frac\zeta 2)\right)
\det S_h(\{\la\}),
\label{intabcd3}
\end{align}
where $\{\la_1,\dots\la_m\}=\{\mu'_{j'_{\mathrm{max}}},\dots,\mu'_{j'_{\mathrm{min}}},
\mu_{j_{\mathrm{min}}},\dots,\mu_{j_{\mathrm{max}}}\}$ and the matrix
elements $m\times m$ matrix $S_h$, 
\[{S_h}_{a b}=\rho_{h,b}(\la_a),\]
are defined as solutions of the following integral equations:
\begin{equation}
\rho_{h,b}(\la)+
\int\limits_{-\Lambda_h}^{\Lambda_h}d\mu K(\la-\mu)\rho_{h,b}(\mu)=\frac 1{2\pi (b-1)!}
\frac{d^b}
{d\la^b} p_0(\la).
\label{hliebb}
\end{equation} 

In the case $\Delta>1$, $h>h_c$ the result has the following form,
\begin{align}
F_m(h,\{\e_j,\e'_j\})=&
\pl_{j=1}^{s'}\left(\int\limits_{-\Lambda_h}^{-\pi/2}+
\int\limits_{-\pi/2-i\zeta}^{\pi/2-i\zeta}+\int\limits_{\pi/2}^{\Lambda_h}\right)d\la_{j} 
\,\pl_{j=s'+1}^m \int\limits_{-\Lambda_h}^{\Lambda_h}d\la_{j} \frac{(-1)^{s'} }{\pl_{k>l}
\sin(\la_k-\la_l-i\zeta)}\times \nonumber\\
\times&\pl_{j\in\mathbf{\a^-}}\left(
\sin^{j-1}(\mu_j-i\frac\zeta 2)\sin^{m-j}(\mu_j+i\frac\zeta 2)\right)\times\nonumber\\
\times&\pl_{j\in\mathbf{\a^+}}\left(
\sin^{j-1}(\mu'_j+3i\frac\zeta 2)\sin^{m-j}(\mu'_j+\frac\zeta 2)\right)
\det S_h(\{\la\}).
\label{intabcd4}
\end{align}
The matrix elements of the matrix $S_h$ are given also by the integral equations
(\ref{hliebb})  with the kernel $K(\la-\mu)$ and bare momentum $p_0(\la)$ corresponding
to the regime $\Delta >1$.

The determinants cannot be calculated explicitly like in the case of zero magnetic field but
however the density functions and hence the determinants can be calculated numerically
from the corresponding integral equations.

\vskip 1cm
{\bf\large Acknowlegement.} We would like to thank A. Izergin and N. Slavnov for useful discussions.


\section*{Appendix A}
\renewcommand{\theequation}{A.\arabic{equation}}
\setcounter{equation}{0}

In this appendix we compute the determinant of the matrix 
$(\tilde{S}_{ij})_{1\le i,j \le m}$, with 
$\tilde{S}_{ij}=\rho(\lambda_i-\beta_j-i\frac{\zeta}{2})$, 
of the density in the massive case. In the domain $\Delta>1$, $\rho$ is an 
elliptic function with periods $\pi$ and $2i\zeta$ ($q=e^{-\zeta}$), 
which can be written in terms of Theta-functions:
\begin{equation}
  \rho(\lambda)=\frac{1}{2\pi}\sul_{n=-\infty}^{+\infty}
                       \frac{e^{2i\lambda n}}{\cosh n\zeta}
               =\frac{1}{2\pi} \pl_{n=1}^{+\infty} 
                \Bigl(\frac{1-q^{2n}}{1+q^{2n}}\Bigr)^2\ 
                \frac{\th_3(\la,q)}{\th_4(\la,q)}.
\end{equation}

To compute the determinant $\det_m \tilde{S}$, let us first consider it as a function $f$
of the variable $\la_1$. $f$ is thus an elliptic function of $\la_1$ of order
$2m$ and with the same periods as $\rho$. An irreducible set of poles is
$\{\beta_1,\dots,\beta_m,
\beta_1+i\zeta,\dots,\beta_m+i\zeta\}$, and 
$\la_2,\dots,\la_m$ are obviously zeros of $f$. Note that, for any $x$,
$f(x+i\zeta)=-f(x)$, hence $\la_2+i\zeta,\dots,\la_m+i\zeta$ are also
zeros of $f$. Up to congruence, there remains also only two other zeros
which differ by $i\zeta$, say $x_0$ and $x_0 +i\zeta$. Since the sum of
zeros of an elliptic function is congruent to the sum of its poles, 
it follows that $x_0$ is congruent either to 
$\sul_{i=1}^m\beta_i-\sul_{i=2}^m\la_i$ or to
$\sul_{i=1}^m\beta_i-\sul_{i=2}^m\la_i-\frac{\pi}{2}$.
Actually it is congruent to the second expression, as it will be shown latter.

Let us now consider the function
\begin{equation*}
  g(\la_1)=\frac{\pl_{j=2}^m \th_1(\la_1-\la_j)}
             {\pl_{j=1}^m \th_1(\la_1-\beta_j)}\ 
             \th_2\biggl(\sul_{j=1}^m(\la_j-\beta_j)\biggr).
\end{equation*}
$g$ is an elliptic function with periods $\pi$ and $2i\zeta$, of order $2m$,
which has the same poles and zeros as $f$. By Liouville's theorem, $f/g$ is a 
constant. Note at this stage that if $x_0$ was taken to be congruent to
$\sul_{i=1}^m\beta_i-\sul_{i=2}^m\la_i$, a similar argument
would have lead to the fact that $f$ should be equal, up to a multiplicative
constant, to,
\begin{equation*}
  \frac{\pl_{j=2}^m \th_1(\la_1-\la_j)}
             {\pl_{j=1}^m \th_1(\la_1-\beta_j)}\ 
             \th_1\biggl(\sul_{j=1}^m(\la_j-\beta_j)\biggr),
\end{equation*}
which is obviously not true because the periods do not coincide.

The same procedure for the variables $\la_2,\dots,\la_m$, and similarly 
$\beta_1,\dots,\beta_m$, leads to the 
following formula for $\det_m \tilde{S}$:
\begin{equation}\label{eq:det-th}
  \det_m \tilde{S} = C_m\, \frac{\pl_{j<k} \th_1(\la_j-\la_k)\ \th_1(\beta_k-\beta_j)}
             {\pl_{j,k=1}^m \th_1 (\la_j-\beta_k)}\ 
              \th_2\biggl(\sul_{j=1}^m(\la_j-\beta_j)\biggr), 
\end{equation} 
where $C_m$ is a constant which does not depend on 
$\la_i,\ \beta_j,\ 1\le i,j \le m$.

A recursion relation for $C_m$ can be obtained by taking the residue of the 
two members of~\eqref{eq:det-th} at the pole $\la_m=\beta_m$:
\begin{equation*}
 C_m=i\frac{q^{\frac{1}{4}}}{\pi} \pl_{n=1}^{\infty} (1-q^{2n})^3\ C_{m-1}.
\end{equation*}
The determination of $C_1$ is straitforward,
\begin{equation*}
 C_1=\frac{i}{2\pi} \pl_{n=1}^{\infty} \Bigl(\frac{1-q^{2n}}{1+q^{2n}}\Bigr)^2,
\end{equation*}
and thus we obtain for the constant $C_m$:
\begin{equation*}
 C_m=\Bigl(\frac{i}{2\pi}\Bigr)^m 
      \pl_{n=1}^{\infty} \Bigl(\frac{1-q^{2n}}{1+q^{2n}}\Bigr)^2
      \Big[2q^{1/4} \pl_{n=1}^{\infty} (1-q^{2n})^3\Big]^{m-1}. 
\end{equation*}
%



\end{document}